\documentclass[12pt]{article}
\usepackage{epsfig, epsf, graphicx, url, bm, natbib, arydshln, subcaption, float}
\usepackage[hidelinks]{hyperref}
\hypersetup{colorlinks = true, urlcolor = blue, linkcolor = blue, citecolor = red}
\usepackage{pstricks, pst-node, psfrag}

\usepackage{amssymb,amsmath}
\usepackage{verbatim,enumerate}
\usepackage{rotating, lscape}
\usepackage{setspace}
\usepackage{multirow}
\def\log{\hbox{log}}

\def\dfrac#1#2{{\displaystyle{#1\over#2}}}

\def\boxit#1{\vbox{\hrule\hbox{\vrule\kern6pt
          \vbox{\kern6pt#1\kern6pt}\kern6pt\vrule}\hrule}}

\def\trace{\hbox{trace}}

\def\bse{\begin{eqnarray*}}
\def\ese{\end{eqnarray*}}
\def\be{\begin{eqnarray}}
\def\ee{\end{eqnarray}}
\def\bq{\begin{equation}}
\def\eq{\end{equation}}
\def\bse{\begin{eqnarray*}}
\def\ese{\end{eqnarray*}}

\newcommand{\bI}{\mathbf{I}}
\newcommand{\bX}{\mathbf{X}}

\newcommand{\bb}{\mathbf{b}}
\newcommand{\bB}{\mathbf{B}}

\newcommand{\bu}{\mathbf{u}}

\newcommand{\balpha}{\boldsymbol{\alpha}}

\newcommand{\btheta}{\boldsymbol{\theta}}
\newcommand{\bphi}{\boldsymbol{\phi}}
\newcommand{\bpsi}{\boldsymbol{\psi}}

\newcommand{\bvarphi}{\boldsymbol{\varphi}}

\newcommand{\bi}{\begin{itemize}}
\newcommand{\ei}{\end{itemize}}

\newcommand\tr{{\mathrm{tr}}}

\newcommand{\bA}{\mathbf{A}}

\newcommand{\bU}{\mathbf{U}}

\newcommand{\bR}{\mathbf{R}}

\newcommand{\bomega}{\bm{\sigma}}
\newcommand{\bTheta}{\bm{\Theta}}

\newcommand{\bPsi}{\mathbf{\Psi}}

\def\bL {\mbox{\boldmath $L$}}

\def\log{\hbox{log}}

\def\boxit#1{\vbox{\hrule\hbox{\vrule\kern6pt
          \vbox{\kern6pt#1\kern6pt}\kern6pt\vrule}\hrule}}

\def\trace{\hbox{trace}}

\def\beq{\begin{equation}}
\def\eeq{\end{equation}}
\def\beqa{\begin{eqnarray}}
\def\eeqa{\end{eqnarray}}
\def\beqs{\begin{equation*}}
\def\eeqs{\end{equation*}}
\def\beqas{\begin{eqnarray*}}
\def\eeqas{\end{eqnarray*}}

\def\bX{{\ensuremath\mathbf{X}}}

\def\dfrac#1#2{\displaystyle{{#1}\over{#2}}}
\def\boxit#1{\vbox{\hrule\hbox{\vrule\kern6pt
          \vbox{\kern6pt#1\kern6pt}\kern6pt\vrule}\hrule}}

\def\bA{{\bf A}}
\def\bX{{\bf X}}

\def\bU{{\bf U}}

\def\bu{{\bf u}}

\def\bb{{\bf b}}

\def\balpha{{\boldsymbol \alpha}}

\def\btheta{{\boldsymbol \theta}}
\def\bomega{{\boldsymbol \omega}}
\def\bphi{{\boldsymbol \phi}}

\def\bTheta{{\boldsymbol \Theta}}

\def\bR{\mathbf{R}}

\def\bX{\mathbb{X}}

\def\A{\mbox{A}}

\def\bX{\mathbb{X}}

\def\A{\mbox{A}}

\def\bB{{\bf B}}
\def\bD{{\bf D}}
\def\bI{{\bf I}}
\def\bX{{\bf X}}

\def\bR{{\bf R}}
\def\bU{{\bf U}}

\def\tr{{\mathrm{tr}}}

\begin{document}

\thispagestyle{empty} \baselineskip=28pt \vskip 5mm
\begin{center} {\Large{\bf Nonparametric collective spectral density estimation with an application to clustering the brain signals}}
\end{center}

\baselineskip=12pt \vskip 5mm

\begin{center}\large
Mehdi Maadooliat\footnote[1]{\baselineskip=10pt Department of Mathematics, Statistics and Computer Science, Marquette University, Milwaukee, WI 53201-1881 USA. E-mail: mehdi@mscs.mu.edu}, Ying Sun\footnote[2]{\baselineskip=10pt CEMSE Division, King Abdullah University of Science and Technology, Thuwal 23955-6900 Saudi Arabia. E-mail: ying.sun@kaust.edu.sa} and Tianbo Chen$^2$
\end{center}

\baselineskip=16pt \vskip 1mm \centerline{\today} \vskip 8mm

\begin{center}
{\large{\bf Abstract}}
\end{center}
In this paper, we develop a method for the simultaneous estimation of spectral density functions (SDFs) for a collection of stationary time series that share some common features. Due to the similarities among the SDFs, the log-SDF can be represented using a common set of basis functions. The basis shared by the collection of the log-SDFs is estimated as a low-dimensional manifold of a large space spanned by a pre-specified rich basis. A collective estimation approach pools information and borrows strength across the SDFs to achieve better estimation efficiency. Also, each estimated spectral density has a concise representation using the coefficients of the basis expansion, and these coefficients can be used for visualization, clustering, and classification purposes. The Whittle pseudo-maximum likelihood approach is used to fit the model and an alternating blockwise Newton-type algorithm is developed for the computation. A web-based \href{https://ncsde.shinyapps.io/NCSDE}{shiny App} found at ``\href{https://ncsde.shinyapps.io/NCSDE}{https://ncsde.shinyapps.io/NCSDE}'' is developed for visualization, training and learning the SDFs collectively using the proposed technique. Finally, we apply our method to cluster similar brain signals recorded by the electroencephalogram for identifying synchronized brain regions according to their spectral densities.

\baselineskip=17pt

\begin{doublespace}
\par\vfill
\noindent{\bf MSC 2010 subject classifications:} Primary 62G05, 62M15; secondary 62P10.\\
\noindent{\bf Keywords and phrases:} Collective estimation; nonparametric estimation; roughness penalty; time series clustering; spectral density function; Whittle likelihood\\
\noindent{\bf Short title:} Nonparametric collective spectral density estimation 
\end{doublespace}

\clearpage\pagebreak\newpage \pagenumbering{arabic}
\baselineskip=26.5pt

\begin{doublespace}

\section{Introduction}\label{sec:intro}

The spectral density plays an important role in time series analysis in frequency domain. We suppose that $X_t$  represents a zero-mean, weakly stationary time series with the autocovariance function $\gamma(h)$ defined as $\gamma(h)=E(X_{t}X_{t+h}),$ $h=0,\pm 1,\pm 2, \ldots.$ If the autocovariance function is absolutely summable,~i.e., $\sum_{h=-\infty}^{\infty}|\gamma(h)|<\infty,$
then the autocovariance sequence $\gamma(h)$ has the spectral representation
$$\gamma(h) = \displaystyle\int_{-1/2}^{1/2}{f(\omega)}e^{2\pi i\omega h}d\omega,$$
where $f(\omega)$ is the spectral density of $X_t$, which has the 
inverse Fourier representation 
$$ f(\omega)=\sum_{h=-\infty}^{\infty}\gamma(h)e^{-2 \pi i \omega h}, \quad -1/2\leq \omega \leq 1/2.$$
Given the time series $\{x_t$, $t=1,\ldots,n\}$, when ${\gamma}(h)$ is replaced by the sample covariance $\hat{\gamma}(h)$, the periodogram is defined as
$${I_n(\omega_j)}=\displaystyle\sum_{h=-(n-1)}^{n+1}\hat{\gamma}(h)e^{-2\pi i\omega_j h},\quad j=0,1,...,n-1,$$
which can be calculated as $I_n(\omega_j) = |d(\omega_j)|^2$, where $\displaystyle d(\omega_j) = n^{-1/2}\sum_{t=1}^n x_t e^{-2\pi i \omega_j t}$
is the discrete Fourier transform of $x_t$ at the fundamental frequencies $\omega_j=j/n$.

However, the raw periodogram is not a consistent estimator for the spectral density of a stationary random process. One classical method to obtain a consistent estimator is to smooth the periodogram across frequencies. \citet{yuen1979smoothed} analyzed the performance of three methods of periodogram smoothing for spectrum estimation. \citet{wahba1980automatic} developed an objective optimum smoothing procedure for estimating the log-spectral density using the spline to smooth the log-periodogram. A discrete spectral average estimator and lag window estimators were introduced in \citet{brockwell2013time}. Both of the two methods are consistent. \citet{brillinger2001time} introduced periodogram kernel smoothing. One critical issue in periodogram smoothing is span selection. \citet{lee1997simple} proposed a span selector based on unbiased risk estimation. \citet{ombao2001simple} proposed using the Gamma-deviance generalized cross-validation (Gamma GCV), and \citet{lee2001stabilized} proposed a bandwidth selection based on coupling of the so-called plug-in and the unbiased risk estimation ideas.

Another popular method for spectral density estimation is based on likelihoods. For example, \citet{capon1983maximum} used the maximum likelihood method to estimate the spectral density of signals with noise. In \citet{guler2001ar}, the spectral density of brain signal data was analyzed using the maximum likelihood. The well-known Whittle likelihood was developed for time series analysis in \citet{whittle1953estimation}, \citet{whittle1954some}, and \citet{whittle1962gaussian}. \citet{pawitan1994nonparametric} proposed using a penalized Whittle likelihood to estimate spectral density functions. \citet{chow1985sieve} developed a penalized likelihood-type method for the nonparametric estimation of the spectral density of Gaussian processes. \citet{fan1998automatic} used local polynomial techniques to fit the Whittle likelihood for spectral density estimation. We suppose there is a time series $\bX$ of length $n$ from a mean-zero stationary Gaussian process with a parametric covariance function $\gamma(h;\btheta)$, where $\btheta$ is the unknown parameter. If we let $\Gamma_{n,\btheta}$ be the covariance matrix of the random vector $\bX$, then the likelihood function is
$$L(\btheta)=\dfrac{1}{(2\pi)^{n/2}|{\Gamma_{n,\btheta}}|^{1/2}}\exp\left(-\dfrac{1}{2}\bX^\top{\Gamma}^{-1}_{n,\btheta}\bX\right),$$
and $\ell(\btheta) = -2\times\log L(\btheta)$ is the log-likelihood,
$$\ell(\btheta) \propto \log(|{\Gamma_{n,\btheta}}|) + \tr\left(\bX\bX^\top{\Gamma}^{-1}_{{n,\btheta}}\right).$$ 
\citet{Whittle1954} proposed an approximation of the above log-likelihood function, known as the Whittle likelihood approximation:
\be \label{Whittle.app}
\ell_W(\btheta) = n\int_{-1/2}^{1/2}\left\{\log({f_\btheta(\omega)}) + I_n(\omega) {f_\btheta(\omega)}^{-1}\right\}d\omega,
\ee
where ${f_\btheta(\omega)}$ is the spectral density function. For a discrete frequency range, the Whittle approximation \eqref{Whittle.app} can be written as
$$\ell_W(\btheta) \stackrel{.}{=} n\sum_{-1/2<\omega_j<1/2}\Bigl\{\log({f_\btheta(\omega_j)}) + {I_n(\omega_j)} {f_\btheta(\omega_j)}^{-1}\Bigr\}.$$
Using the fast Fourier transform (FFT), $\ell_W(\btheta)$ can be obtained efficiently with only $O\left(n\log_2 n\right)$ operations.


We apply our method to the neuroscience study of functional connectivity in the brain with electroencephalgram (EEG) data.
The EEG is a method of monitoring spontaneous electrical activity in the brain over a period of time. EEGs are typically recorded from multiple electrodes, referred as EEG channels, placed on the scalp. The clustered EEG channels are useful for understanding how different brain regions communicate with each other. To identify synchronized EEG channels that show similar oscillations or waveforms, many time series clustering algorithms have been developed to cluster similar smoothed periodograms estimated from the EEG channels separately \citep[see][for some examples]{Hennig:2015}.

In this paper, we propose collectively estimating multiple spectral densities that share common features. The collective estimation was first proposed by \citet{Maadooliat2016} for the estimation of probability density functions in protein structure analysis. For multiple stationary time series that may share similar spectral density functions, we propose a nonparametric spectral density estimation approach that minimizes a penalized Whittle likelihood.  The dimension is reduced by representing the raw periodograms on a set of common basis functions while allowing each one to differ by introducing random effects into the model. The collectively estimated spectral densities can then be used for clustering purposes. Collective estimation is a novel approach for estimating multiple spectral densities that share common features in a statistically more efficient way. 

The rest of the paper is organized as follows. Section~\ref{sec:meth} presents the core of the proposed method. Subsection \ref{sec:estimation} provides the blockwise Newton-Raphson algorithm, and subsections \ref{sec:tuning}-\ref{sec:identifiability} provide the implementation details. Section \ref{sec:sim} reports the simulation results to illustrate the proposed collective estimation approach and to compare it with competitive non-collective estimation approaches. The application to EEG data is given in Section \ref{sec:app}. Section \ref{sec:end} concludes the paper.

\section{Methodology}\label{sec:meth}

We consider a collection of stationary time series, $\bX_i$s, where $i=1, \cdots, m$\quad and \quad $\bX_i^\top = (X_{i1}, X_{i2}, \cdots, X_{in})$ with the associated spectral density $f_i$. We want to estimate the spectral density functions together. The rationale of this collective spectral density estimation approach is to improve the efficiency of the estimates by using a shared basis to obtain the SDFs.

We assume that each log-spectral density function can be represented by a linear combination of a common set of basis functions $\{\phi_k(\omega), k=1, \cdots, K\}$, and that each has its own set of coefficients. More specifically, we assume that $\log\{f_i(\omega)\} = u_i(\omega)$, with
\beq \label{eq:adapt-basis}
u_i(\omega)=\sum_{k=1}^K \phi_k(\omega)\alpha_{ik}, \qquad i =1, \dots, m.
\eeq
Equivalently, the spectral density functions can be written as 
\beq\label{eq:adapt-basis2}
f_i(\omega) = \exp{u_i(\omega)} = \exp \Biggl\{\sum_{k=1}^K \phi_k(\omega)\alpha_{ik}\Biggr\},
\quad i =1, \dots, m.
\eeq
For identifiability, we require that $1, \phi_k,\ k=1, \dots, K$, to be linearly independent. We would like $K$ to be a small number so that the number of parameters to be estimated remains manageable even when we estimate a large number of spectral densities (i.e., when $m$ is large).

In our setting, the basis functions are not prespecified and need to be determined from the data. To this end, we suppose that these basis functions fall in a low-dimensional subspace of a function space spanned by a rich family of fixed basis functions, $\{b_\ell(\omega), \ell=1, \cdots, L\}$ ($L \gg K$), such that
\beq \label{eq:fixed-basis}
\phi_k(\omega) = \sum_{\ell=1}^L b_\ell(\omega) \theta_{{\ell}k}, \qquad k = 1\dots, K.
\eeq
For identifiability, we require that $1,\ b_\ell,\ \ell=1, \dots, L$, to be linearly independent. A large enough $L$ ensures the necessary flexibility to represent the unknown spectral densities. For univariate cases, the fixed basis can be the monomials, the B-splines, or the Fourier basis. Bivariate splines can be used as the fixed basis functions for bivariate spectral densities.

To simplify the presentation, we now introduce some vectors and matrices to denote the quantities of interest: $\bphi(\omega) = ( \phi_1(\omega), \cdots, \phi_K(\omega) )^\top$, $\balpha_i = ( \alpha_{i1}, \cdots, \alpha_{iK} )^\top$, $ \bb(\omega) = ( b_1(\omega), \cdots, b_L(\omega) )^\top$,  $\btheta_k = ( \theta_{1k}, \cdots, \theta_{Lk} )^\top$, and $\bTheta =  ( \btheta_1, \cdots, \btheta_K )$.  Then, from \eqref{eq:adapt-basis} and \eqref{eq:fixed-basis}, we can rewrite $u_i(\omega)$ in the vector-matrix form as
\beq \label{model:omega}
u_i(\omega)= \bphi(\omega)^\top \balpha_i = \bb(\omega)^\top \bTheta \balpha_i, \qquad i =1, \dots, m.
\eeq
One may combine the equations above for $i =1, \dots, m$ into a matrix form by evaluating the log-SDFs over the discrete frequencies $\bomega=(\omega_1,\cdots,\omega_{\tilde{n}})^\top$. The $m$ equations given in \eqref{model:omega} can be written as $\bU=\bB\bTheta\bA^\top$, where $\bU=\left(u_{1}(\bomega),\cdots,u_{m}(\bomega)\right)$ is an $\tilde{n}\times m$ matrix that represents the log-SDFs, $\bB=\left(\bb(\omega_1),\cdots,\bb(\omega_{\tilde{n}})\right)^\top$ is an $\tilde{n}\times L$ matrix that represents the rich basis functions at the discrete frequencies $\bomega$, and $\bA = (\balpha_1, \dots, \balpha_m)^\top$.  The unknown parameters can then be collectively written as the pair $(\bTheta, \bA)$.  There is an identifiability issue caused by the non-uniqueness of the parametrization of $(\bTheta, \bA)$. This issue can be resolved by introducing some restrictions on the parameterization; see subsection \ref{sec:identifiability}.

We could have used the fixed basis $\{b_\ell(\omega), \ell=1, \cdots, L\}$  in \eqref{eq:adapt-basis} and \eqref{eq:adapt-basis2}; however, that would be either too restrictive (if $L$ is small) or produce a large  number of parameters (if $L$ is large). Alternatively, if we were to model the individual density functions separately using the fixed basis $\{b_\ell(\omega), \ell=1, \cdots, L\}$, then we would write
\beq \label{model:omega2}
u_i(\omega) = \bb(\omega)^\top \bpsi_i, \qquad i =1, \dots, m.
\eeq
We let $\bPsi = (\bpsi_1, \dots, \bpsi_m)$ be the  $L \times m$ matrix of coefficients from the basis expansions given in \eqref{model:omega2}. By comparing \eqref{model:omega} and \eqref{model:omega2}, we obtain $\bPsi=\bTheta\bA^\top$, which is a rank $K$ matrix.  Thus, the collective modeling approach introduces a low-rank structure to the coefficient matrix in the basis expansion of the log-spectral densities. This dimensionality reduction allows us to significantly reduce the number of parameters to be estimated and, thus, gain estimation efficiency.

\subsection{Estimation using penalized Whittle likelihood}
\label{sec:estimation}

The periodogram $I_{i,n}$ is a rough estimate of the spectral density, $f_i$, associated with the time series $\bX_i$ observed over $n$ time points:
$$I_{i,n}(\omega)={\Bigl|\dfrac{1}{n}\sum_{t=1}^n X_{it}e^{-2\pi it\omega}\Bigr|^2}.$$
We consider the periodogram at the Fourier frequencies $\bm{\omega}=2\pi\bm{J}/n$ for $\bm{J}=\{-\lfloor(n-1)/2\rfloor,\cdots,n-\lfloor n/2\rfloor\}.$

The Whittle likelihood for estimating $m$ spectral densities has the following form:

\beq\label{eq:loglik}
\ell_W(\bTheta, \bA) = \sum_{i=1}^m \sum_{{j}\in\bm{J}} \Biggl\{u_i(\omega_{j}) + I_{i,n}(\omega_{j})\exp\bigl[-u_i(\omega_{j})\bigr]\Biggr\},
\eeq
where $u_i(\omega)$ are defined in \eqref{model:omega}.  It is concave in $\balpha_i$ when other parameters are fixed and also concave in $\btheta_k$ when other parameters are fixed. Applying the roughness penalty approach of function estimation \citep{Green1994}, we estimate the model parameters by minimizing the penalized likelihood criterion
\begin{equation}
- 2\, \ell_W(\bTheta, \bA) + \lambda \sum_{k=1}^K\, {\sf PEN}(\phi_k),\label{eq:penLik}
\end{equation}
where ${\sf PEN}(\phi_k)$ is a roughness penalty function that regularizes the estimated basis function $\phi_k$ to ensure that it is a smooth function, and $\lambda >0$ is a penalty parameter. The penalty function can be written in a quadratic form as
\begin{equation}\label{eq:pen}
\sum_{k=1}^K\, {\sf PEN}(\phi_k) = \sum_{i=k}^K \btheta_k^\top \bR \btheta_k = \tr\{ \bTheta^\top \bR \bTheta\}.
\end{equation}
Two choices for the penalty matrices ($\bR_1$ and $\bR_2$), based on the second derivative of the basis functions and the difference operator, are given in subsection \ref{penalty}.

We use an alternating blockwise Newton-Raphson algorithm to minimize the penalized Whittle likelihood approximation. Our algorithm cycles through updating $\balpha_i$ for $i=1, \dots,m$ and $\btheta_k$ for $k=1, \dots, K$ until convergence. Following the usual step-halving strategy for the Newton-Raphson iteration, the updating formulas are

\beqa\label{eq:update-alpha}
\balpha_i^{new} & = & \balpha_i^{old} - \tau \, \biggl[\frac{\partial^2}{\partial \balpha_i \partial \balpha_i^\top}
\{\ell_W(\bTheta, \bA) \}\biggr]^{-1} \biggl[\frac{\partial}{\partial \balpha_i} \{ \ell_W(\bTheta, \bA)\}\biggr]
\bigg\vert_{\bTheta=\bTheta^{old}, \bA=\bA^{old}} \nonumber \\
& = & \balpha_i^{old} - \tau\biggl[\bTheta^\top\sum_{j}\Bigl\{ \bb(\omega_{j}) I_{i,n}(\omega_{j})\exp\bigl[-u_i(\omega_{j})\bigr] \bb(\omega_{j})^\top \Bigr\}\bTheta\biggr]^{-1} \times\nonumber\\ && \quad
\biggl[\bTheta^\top\sum_{j}\Bigl\{ \bb(\omega_{j}) - \bb(\omega_{j}) I_{i,n}(\omega_{j})\exp\bigl[-u_i(\omega_{j})\bigr] \Bigr\}\biggr]\bigg\vert_{\bTheta=\bTheta^{old}, \bA=\bA^{old}}
\eeqa

and

\beqa\label{eq:update-theta}
\btheta_k^{new} \ = \btheta_k^{old} - \tau \, \biggl[\frac{\partial^2}{\partial \btheta_k \partial \btheta_k^\top}
\{\ell_W(\bTheta, \bA)  \} - \lambda \bR \biggr]^{-1}\times\qquad\qquad\qquad\qquad\qquad\qquad \nonumber\\
\qquad\qquad\qquad\biggl[\frac{\partial}{\partial \btheta_k} \{ \ell_W(\bTheta, \bA) \}
- \lambda \bR \btheta_k\biggr]\bigg\vert_{\bTheta=\bTheta^{old}, \bA=\bA^{old}}, \nonumber \\
 = \btheta_k^{old} - \tau\biggl[\sum_{i=1}^m\alpha_{ik}^2\sum_{j}\Bigl\{ \bb(\omega_{j}) I_{i,n}(\omega_{j})\exp\bigl[-u_i(\omega_{j})\bigr] \bb(\omega_{j})^\top \Bigr\} - \lambda \bR\biggr]^{-1} \times\nonumber\qquad\\  \qquad \biggl[\sum_{i=1}^m\alpha_{ik}\sum_{j}\Bigl\{ \bb(\omega_{j}) - \bb(\omega_{j}) I_{i,n}(\omega_{j})\exp\bigl[-u_i(\omega_{j})\bigr] \Bigr\} - \lambda \bR \btheta_k\biggr]\bigg\vert_{\bTheta=\bTheta^{old}, \bA=\bA^{old}}
\eeqa

where $\tau$ is the first result from the sequence $\{(1/2)^\delta, \delta=0, 1, \dots\}$ such that the objective function in \eqref{eq:penLik} is reduced. The initial values of the Newton-Raphson iteration can be obtained by projecting the raw spectral density estimates (e.g., peridograms) to the model space of \eqref{eq:adapt-basis2}.

\subsection{Selecting the tuning parameter}\label{sec:tuning}

We may select the penalty parameter by minimizing the AIC \citep{Akaike1974},
\begin{equation}\label{eq:cv-app}
{\rm AIC}(\lambda) = - 2\, \ell_W(\widehat{\bTheta}, \widehat{\bA}) + 2\,{\sf df}(\lambda),
\end{equation}
where $\ell_W(\bTheta, \bA)$ is the log-likelihood defined in \eqref{eq:loglik}, and the degrees of freedom ${\sf df}(\lambda)$ is defined as

\beqa\label{eq:df}
{\sf df}(\lambda) = \sum_{k=1}^K {\rm trace}\biggl\{\biggl[\sum_{i=1}^m\alpha_{ik}^2\sum_{j}\Bigl\{ \bb(\omega_{j}) 
I_{i,n}(\omega_{j})\exp\bigl[-u_i(\omega_{j})\bigr] \bb(\omega_{j})^\top \Bigr\} - \lambda \bR \biggr]^{-1} \times \nonumber \biggr. \nonumber \\ \biggl. 
\biggl[\sum_{i=1}^m\alpha_{ik}^2\sum_{j}\Bigl\{ \bb(\omega_{j}) I_{i,n}(\omega_{j})\exp\bigl[-u_i(\omega_{j})\bigr] \bb(\omega_{j})^\top \Bigr\} \biggr]\biggr\}.\qquad\quad
\eeqa

The parameters in these formulas are replaced by their estimated values. The AIC can be derived as an approximation of the leave-one-out cross-validation \citep{OSullivan1988, Gu2002}.

Selecting the tuning parameter that minimizes the AIC requires training the model for different values of $\lambda$s and then picking the one that minimizes the criterion function, which can be very expensive in time. Instead, we present an alternative procedure that updates the value of the tuning parameter within the Newton-Raphson iterations. This idea has been used in a generalized mixture model to iteratively update the smoothing parameter \citep{Schall1991}. \citet{Schellhase2012} and  \citet{Najibi2017} extended this approach for density estimation. We borrow their formulation, and use the parameter estimates in the $i^{th}$ step to update the tuning parameter, $\hat{\lambda}_{i+1}$, through 
\begin{equation} \label{lambda}
\hat{\lambda}_{i+1}^{-1} = \frac{\trace (\hat\bTheta_i^\top \bR \hat\bTheta_i)}{{\sf df}(\hat\lambda_{i})-(a-1)},
\end{equation}

\noindent where $a$ is the order of the differences  (derivative) used in the penalty matrix $\bR$ (see Section \ref{penalty}). From what we have seen in the implementation of the new procedure, updating the tuning parameter within the Newton-Raphson iterations, on average, does not increase the number of iterations required for convergence. Therefore, the new procedure obtains the final result $p$ times faster than the old procedure, where $p$ is the number of $\lambda$s used in the grid search to minimize the AIC.

\subsection{Choices of penalties}\label{penalty}
\label{sec:penalty}

\subsubsection{Second derivative of the basis}
For the univariate spectral density estimation, if we use the usual squared-second-derivative penalty  ${\sf PEN}(\phi_k) = \int \{\phi_k''(\omega)\}^2 \, dx$ and $\phi_k(\omega) = \bb(\omega)^\top \btheta_k$, then $\bR_1 = \displaystyle\int \ddot{\bb}(\omega) \ddot{\bb}(\omega)^\top\,d\omega$  with $\ddot{\bb}(\omega) = ({b}_1''(\omega), \dots, b_L''(\omega))^\top$.

\subsubsection{Difference operator}
We can also control the roughness of the estimated functions by using the difference penalty \citep{eilers1996} to achieve the appropriate level of smoothness. The variability is controlled through a difference function of order $a$, $\Delta_a$, where $\Delta_1\btheta_k := \btheta_k - \btheta_{k-1}$, and $\Delta_a$ is obtained recursively. For example, the second order difference function,  $\Delta_2$, has the following form: 
$$\Delta_2\btheta_k := \Delta_1\Delta_1\btheta_k = \btheta_k - 2\btheta_{k-1} + \btheta_{k-2}.$$

\noindent We can write the difference functions $\Delta_a$ into a matrix form, $\bL_a$. For example, when $a=1$ we have
 $$ \bL_1= {\left[ {\begin{array}{*{20}{c}}
1&{ - 1}&0& \ldots &0\\    0&1&{ - 1}& \ddots &0\\    \vdots & \ddots & \ddots & \ddots &0\\    0& \cdots &0& 1&{-1}
\end{array}} \right]_{(M-1 \times M)}}$$

\noindent The positive definite penalty matrix used to control the smoothness is denoted as $\bR_2$, and it has the quadratic form $\bR_2=\bL_a^\top\bL_a$. 

In the web-application \href{https://ncsde.shinyapps.io/NCSDE}{NCSDE}, we incorporated both penalties and either one can be used by the user.

\subsection{Number of basis functions (clusters)}
\label{sec:nbasis}

A critical step in dealing with real data is identifying the number of common basis functions (choice of $K$ in the context of NCSDE), which is directly related to the number of clusters, $\tilde{k}$. We identify the number of basis functions (clusters), using the \textit{elbow method} as typically used in clustering analysis and it can be traced back to the work by \citet{Thorndike1953}. First, we obtain $\log(\mathsf{S.Ps}) = \bB\left(\bB^\top \bB\right)^{-1}\bB^\top\log(\bI)$, which is then used as an input to the elbow method, implemented in the R package \textsf{factoextra} \citep{Kassambara2016}, to determine the number of clusters and basis functions. 

In the elbow method, we use hierarchical clustering for partitioning the data, and obtain the total within-cluster sum of squares (\textsf{WSS}) based on different numbers of clusters $\tilde{k}$ (e.g., by varying $\tilde{k}$ from $1$ to $10$). We note that \textsf{WSS} measures the concentration of the clusters and it is desired to be as small as possible. Therefore, the optimal number of clusters can be obtained by plotting \textsf{WSS} against the number of clusters $\tilde{k}$. The location of an elbow (turning point) in this plot is generally considered an indicator of the appropriate number of clusters (see Figures \ref{fig1:fig}(a), \ref{first}(a) and \ref{last}(a) for illustration purposes). From what we have seen in the implementation of NCSDE, a clear-cut elbow is achievable in longer time series ($n\geq 400$).

\subsection{Identifiability of $(\Theta, \A)$} 
\label{sec:identifiability}

The non-uniqueness of the parametrization of $(\bTheta, \bA)$ causes an identifiability issue. Specifically, if $\bU$ is a $K\times K$ orthogonal matrix, then $\bTheta \balpha_i = (\bTheta \bU) (\bU^\top \balpha_i)$. Thus, $\tilde\bTheta = \bTheta \bU$ and $\tilde\balpha_i = \bU^\top \balpha_i$ give the same representation of \eqref{model:omega}. To gain identifiability, we require that $(i)$ $\bTheta^\top \bTheta = \bI$, $(ii)$ $\bA^\top \bA = \bD^2$ is a diagonal matrix, $(iii)$ the columns of $\bA$ are ordered such that the diagonal elements of $\bD^2$ are in  strictly decreasing order, and $(iv)$ the first non-zero element of each column of $\bTheta$ is positive. With such $\bTheta$ and $\bA$, if the diagonal elements of $\bD$ are all different, and we set $\bar{\bA} = \bA \bD^{-1}$ so that $\bar\bA^\top \bar\bA = \bI$, then we have $\bTheta \bA^\top = \bTheta \bD \bar{\bA}^\top$, which is a uniquely defined singular value decomposition (SVD). The desired identifiability of $(\bTheta, \bA)$ then follows from the uniqueness of the SVD.

\section{Simulation study}\label{sec:sim}

We conducted a simulation study to evaluate the proposed nonparametric collective spectral density estimation method and compare it with non-collective spectral density estimation approaches. First, we describe the simulation setup and, in the next subsection, we list the competitive non-collective alternatives.

\subsection{Simulation setup}
In the simulations, we considered autoregressive models of order three, $AR(3)$. In each simulation run, we generated an $n\times m$ matrix $\bX=(\bX_1, \bX_2, \cdots, \bX_m)$, such that $\bX_i^\top = (X_{i1}, X_{i2}, \cdots, X_{in})$. Each $\bX_i$ was a time series of length $n$ from one of the following models:

	$$\left\{\begin{matrix}\mathrm{Model\ I} & AR(3):\ \bvarphi=(0.1,0.5,0.1), & \mathrm{with\ prob.}\ p_1,\\
		\mathrm{Model\ II} & AR(3):\ \bvarphi=(0.1,0.1,0.5), & \mathrm{with\ prob.}\ p_2,\\
		\mathrm{Model\ III} &AR(3):\ \bvarphi=(0.5,0.1,0.1), & \mathrm{with\ prob.}\ p_3.
		\end{matrix}\right.$$
		
\noindent The $AR(3)$ model is given by $X_t = \sum_{i=1}^3 \varphi_i X_{t-i}+ \varepsilon_t$, with the associated SDF
\be
f(\omega) = {\sigma_\varepsilon^2}{\left| 1-\sum_{k=1}^3 \varphi_k e^{-2 \pi i k \omega} \right|^{-2}},
\ee
\noindent where $\sigma_\varepsilon^2 = Var(\varepsilon_t)$ \citep[see][for details]{vonStorch2001}. We set $n=(100,200,400)$, $m=(6,15,30)$, and $p_k=\dfrac{1}{3}$, for $k=1,2,3.$ Therefore, we had nine different pairs of $(n,m)$ and, for each pair we ran the simulation {$N=100$} times.

\subsection{Competitive approaches}\label{comp.approach}
In each simulation run we used the $n\times m$ data matrix $\bX$ as an input to obtain estimates of $m$ SDFs ($\hat{\bf f}$: $\tilde{n}\times m$ matrix) from the following six methods:
\bi
\item{Periodograms (\textsf{Ps}):}
	\subitem We used the periodogram $\bI$ as a rough estimate. For each run, the $i^{th}$ column of the $\tilde{n}\times m$ matrix $\bI$ was a vector of size $\tilde{n}$ obtained via
	$$I_{i,n}(\bomega)={\Bigl|\dfrac{1}{n}\sum_{t=1}^n X_{it}e^{-2\pi it\bomega}\Bigr|^2},$$
evaluated at the discrete frequencies $\bomega=\left(\omega_1,\cdots,\omega_{\tilde{n}}\right).$
\item{Smoothed Periodograms (\textsf{S.Ps}):}
	\subitem The next estimate was obtained by smoothing $\bI$ using the rich set of basis functions, $\mathsf{S.Ps} = \exp \Bigl[ \bB \bigl( \bB^\top \bB \bigr)^{-1} \bB^\top \log( \bI ) \Bigr]$.
\item{tSVD Periodograms (\textsf{tSVD.Ps}):}
	\subitem We used the truncated SVD to obtain the rank $K$ approximation of the smoothed periodograms $\left(\bB^\top \bB\right)^{-1}\bB^\top\log(\bI)$, called $\bTheta_1\bA_1^\top$. This was referred to in \citet{Stewart1993} as the approximation theorem (a.k.a. the Eckart-Young theorem). Therefore, the third estimate, ``tSVD Periodograms'', is {$\mathsf{tSVD.Ps}=\exp\left[\bB\bTheta_1\bA_1^\top\right]$.}
\item{Separate Estimations (\textsf{NSDE}):}
	\subitem We avoided collective estimation and obtained the nonparametric spectral density estimates (NSDE) by maximizing the Whittle likelihood using the rich family of basis functions $\bB$ {\it separately}. From \eqref{model:omega2}, $u_i(\bomega)=\bB\bpsi_i$. Therefore, our fourth estimate is {$\mathsf{NSDE}=\exp\left[\bB\bPsi\right]$}, where $\bPsi = (\bpsi_1, \dots, \bpsi_m)$.
\item{tSVD Separate Estimations (\textsf{tSVD.NSDE}):}
	\subitem Having obtained $\bPsi$ by maximizing the Whittle likelihood separately, we used the truncated SVD to obtain the rank $K$ approximation of $\bPsi$ called $\bTheta_s\bA_s^\top$. The fifth estimate is {$\mathsf{tSVD.NSDE}=\exp\left[\bB\bTheta_s\bA_s^\top\right]$}.
\item{Collective Estimations (\textsf{NCSDE}):}
	\subitem Finally, we used the proposed method from Section \ref{sec:meth} to obtain {$\mathsf{NCSDE}=\exp\left[\bB\widehat{\bTheta}\widehat{\bA}^\top\right]$}.
\end{itemize}

\subsection{Measures of quality}\label{rand.index}
\subsubsection{Adjusted Rand similarity coefficient}
The adjusted Rand similarity coefficient (adjusted Rand index) was used to compare the (dis-)similarity between two clustering results. It is defined as follows \citep{Vinh:2009}:
$$\mathsf{ARI} = \frac{ \sum_{i=0}^1\sum_{j=0}^1 \binom{n_{ij}}{2} - \Bigl[\sum_{i} \binom{n_{i\cdot}}{2} + \sum_{j} \binom{n_{\cdot j}}{2}\Bigr]  / \binom{m}{2} }{ \frac{1}{2} \Bigl[\sum_{i} \binom{n_{i\cdot}}{2} + \sum_{j} \binom{n_{\cdot j}}{2}\Bigr] - \Bigl[\sum_{i} \binom{n_{i\cdot}}{2} + \sum_{j} \binom{n_{\cdot j}}{2}\Bigr]  / \binom{m}{2} }.$$

To calculate the adjusted Rand index, we computed the $2\times 2$ contingency table, which consists of the following four cells:
		\bi
		\item $n_{11}$: the number of observation pairs where both observations are comembers in both clusterings.
		\item $n_{10}$: the number of observation pairs where the observations are comembers in the first clustering but not the second.
		\item $n_{01}$: the number of observation pairs where the observations are comembers in the second clustering but not the first.
		\item $n_{00}$: the number of observation pairs where no pairs are comembers in either clustering result.
		\ei
Furthermore $n_{i\cdot}$ and $n_{\cdot j}$ are defined as $n_{i0}+n_{i1}$ and $n_{0j}+n_{1j}$, respectively. Note that the $\mathsf{ARI}$ ranges from $0$ to $1$, with $0$ indicating that the two clusters do not agree on any pairs and $1$ indicating that the clusters are exactly the same. In subsection \ref{sim.results}, we present and compare the clustering results from the approaches given in subsection \ref{comp.approach}, considering the original labels from the simulation as the gold standard. We did not use the class labels when applying the clustering algorithms; we only use the class labels to evaluate the clustering results.
\subsubsection{Canonical angle}
The canonical angle between the column spaces of $\bu=\log(\bf f)$ and $\hat{\bu}=\log(\hat{\bf f})$ is defined as the maximum angle between any two vectors from the two spaces. Mathematically, it can be computed as
$\mathsf{angle}=\cos^{-1}(\rho)\times180/\pi$, where $\rho$ is the minimum singular value of the matrix $\mathbf{Q}_{\hat{\bu}}^\top\mathbf{Q}_{\bu}$,
and $\mathbf{Q}_{\hat{\bu}}$ and $\mathbf{Q}_{\bu}$ are orthonormal matrices obtained by the QR decomposition of matrices $\hat{\bu}$ and $\bu$, respectively \citep{Golub2013}. Smaller values of the {\sf angle} indicate better estimations of the SDF weight functions.

\subsection{Simulation results}\label{sim.results}
\begin{figure*}[!th]
\includegraphics[width=\textwidth]{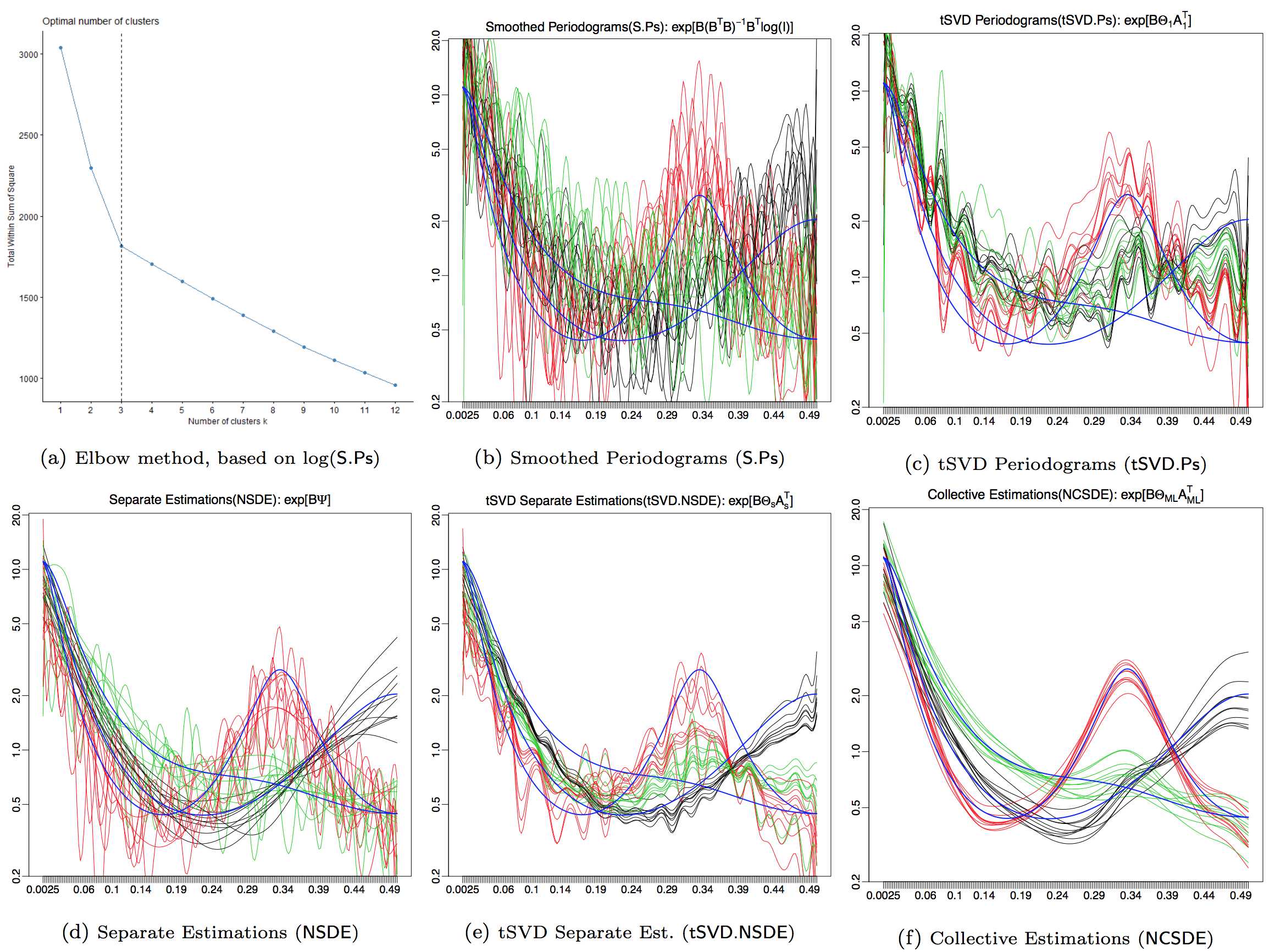}
\caption{For a randomly selected simulation run with $m=30$ and $n=400$. (a) The elbow method suggests picking $\tilde{k}=3$ clusters. (b)-(f) Comparison of results from five different estimates of SDFs (\textsf{S.Ps}, \textsf{tSVD.Ps}, \textsf{NSDE}, \textsf{tSVD.NSDE}, and \textsf{NCSDE}) vs. true SDFs (blue).}
\label{fig1:fig}
\end{figure*}

In each simulation run, we used the approaches given in subsection \ref{comp.approach} for nine different pairs of ($m,n$) to obtain matrices of size $\tilde{n}\times m$ that represent the associated estimates of $m$ SDFs for each approach. Figure \ref{fig1:fig}(a) illustrates how to use the elbow method to obtain the number of clusters ($\tilde{k}=3$), and the remaining subfigures (Figure \ref{fig1:fig}(g)-(f)) provide the results of the five approaches (\textsf{S.Ps}, \textsf{tSVD.Ps}, \textsf{NSDE}, \textsf{tSVD.NSDE}, and \textsf{NCSDE}) when estimating the SDFs for a randomly selected simulation run with $m=30$ and $n=400$. {\sf NCSDE} obtained the results that are the smoothest and the closest to the true SDFs. In order to further verify the effectiveness of {\sf NCSDE} technique, we used the canonical {\sf angle} (details are given in the previous subsection) as a measure of closeness. Table \ref{sim.table.2} shows that $(i)$ the dimension reduction technique clearly improved the efficiency of the results (\textsf{tSVD.Ps} is better than \textsf{Ps}, and \textsf{tSVD.NSDE} is better than \textsf{NSDE}), and $(ii)$ incorporating the dimension reduction technique within the iterative procedure of estimating the SDFs (\textsf{NCSDE}) outperformed the sequential approach of estimating the SDFs {\it separately} and then proceeding with the dimension reduction technique (\textsf{tSVD.NSDE}).

\begin{table*}[!th]
\begin{center}
\caption{{Comparison based on $100$ simulation runs. The mean and standard errors (in parentheses) of the {\sf angle} are reported.}} \label{sim.table.2}
\begin{tabular}{cccccc}
  \hline
  ts-$n$-$m$& \textsf{Ps} & \textsf{tSVD.Ps} & \textsf{NSDE} & \textsf{tSVD.NSDE} & \textsf{NCSDE} \\
  \hline
  ts-100-06 & 84.97(0.38) & 78.47(0.77) & 88.20(0.17) & 68.51(1.74) & 37.50(1.98) \\
  ts-100-15 & 89.13(0.08) & 80.50(0.71) & 90.00(0.00) & 79.10(0.80) & 33.84(2.64) \\
  ts-100-30 & 88.93(0.09) & 80.66(0.73) & 90.00(0.00) & 79.10(0.70) & 24.12(1.31) \\ \hdashline
  ts-200-06 & 84.93(0.43) & 75.40(0.85) & 89.65(0.04) & 59.34(2.14) & 20.36(0.50) \\
  ts-200-15 & 88.29(0.14) & 78.40(0.95) & 90.00(0.00) & 51.74(2.28) & 15.50(0.89) \\
  ts-200-30 & 89.38(0.05) & 80.59(0.69) & 90.00(0.00) & 66.85(1.53) & 11.80(0.32) \\ \hdashline
  ts-400-06 & 85.51(0.36) & 66.16(1.30) & 89.90(0.01) & 43.12(1.72) & 13.91(0.55) \\
  ts-400-15 & 87.66(0.18) & 71.88(1.16) & 90.00(0.00) & 38.41(1.53) & 12.39(0.34) \\
  ts-400-30 & 88.92(0.08) & 73.07(1.18) & 90.00(0.00) & 34.02(1.45) &\ 7.76(0.17) \\ \hline
\end{tabular}
\end{center}
\end{table*}

\begin{figure}[!th]
  \centering  \includegraphics[width=\textwidth]{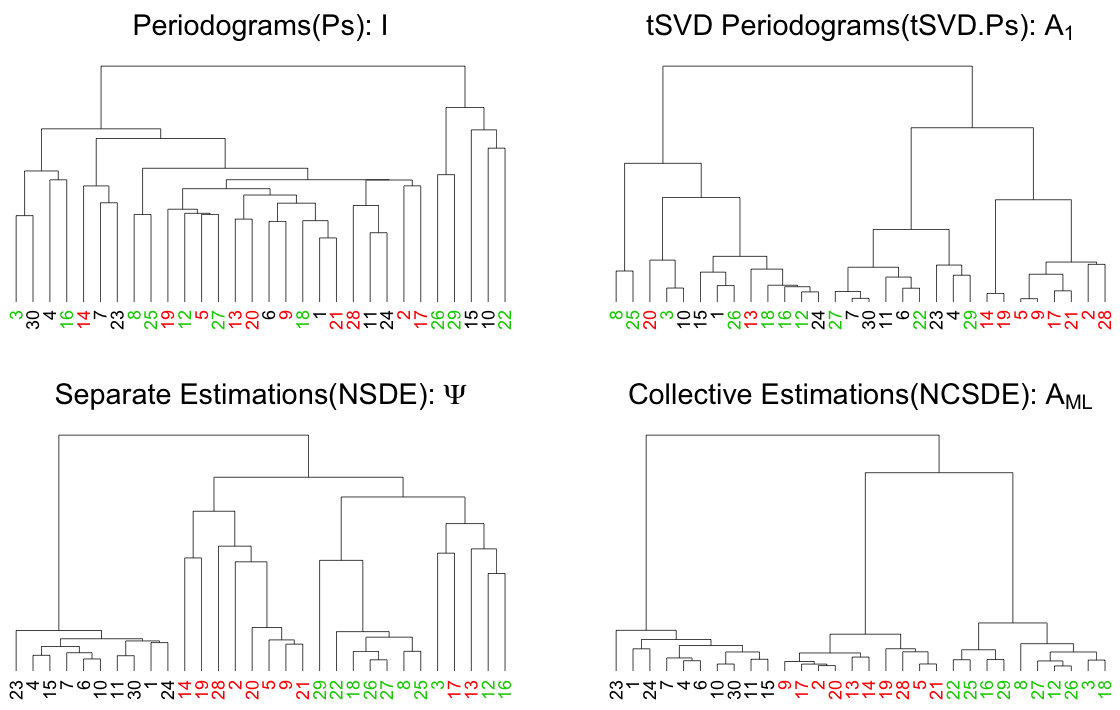}
  \caption{Dendrograms from hierarchical clustering. Comparing four different estimates (\textsf{Ps}, \textsf{tSVD.Ps}, \textsf{NSDE}, and \textsf{NCSDE}) for a randomly selected simulation run with $m=30$ and $n=400$.} \label{fig2:fig}
\end{figure}

The distance matrices obtained from the approaches given in subsection \ref{comp.approach} can be used as an input to the hierarchical clustering algorithm. In our \href{https://ncsde.shinyapps.io/NCSDE}{shiny App}, we implemented the {\sf hclust} function with option {\sf \{method="ward.D2"\}} in the R package {\sf stats} to obtain dendrograms \citep{R2016} (see Figure \ref{fig2:fig} for a comparison of four of the approaches, \textsf{Ps}, \textsf{tSVD.Ps}, \textsf{NSDE}, and \textsf{NCSDE}, on a randomly selected simulation run with $m=30$ and $n=400$). Since we knew the number of clusters and the cluster labels (gold standard) in advance, we cut the dendrograms into the original number of clusters.

To evaluate the performance of the proposed method in discovering the correct label (gold standard), in a more systematic framework, we used the popular external measure, adjusted Rand index, which is commonly used in the clustering evaluation literature and is discussed in subsection \ref{rand.index}. The associated results are given in Table \ref{sim.table.1}. \textsf{NCSDE} clearly outperformed the competing approaches in this clustering task, based on the {adjusted Rand index similarity} ($\mathsf{ARI}$).

\begin{table*}[!th]
\begin{center}
\caption{{Comparison based on $100$ simulation runs. The mean and standard errors (in parentheses) of the {adjusted Rand similarity coefficient} ($\mathsf{ARI}$) are reported.}} \label{sim.table.1}
\begin{tabular}{cccccc}
  \hline
  ts-$n$-$m$ & \textsf{Ps} & \textsf{tSVD.Ps} & \textsf{NSDE} & \textsf{tSVD.NSDE} & \textsf{NCSDE} \\
  \hline
  ts-100-06 & 0.651(0.012) & 0.631(0.010) & 0.686(0.014) & 0.691(0.014) & 0.914(0.012) \\
  ts-100-15 & 0.542(0.007) & 0.559(0.005) & 0.483(0.007) & 0.483(0.007) & 0.952(0.008) \\
  ts-100-30 & 0.526(0.006) & 0.550(0.003) & 0.464(0.005) & 0.464(0.005) & 0.959(0.006) \\ \hdashline
  ts-200-06 & 0.639(0.010) & 0.640(0.011) & 0.742(0.016) & 0.731(0.016) & 0.996(0.003) \\
  ts-200-15 & 0.559(0.008) & 0.570(0.006) & 0.784(0.020) & 0.681(0.022) & 0.994(0.003) \\
  ts-200-30 & 0.526(0.005) & 0.547(0.003) & 0.636(0.018) & 0.567(0.015) & 0.998(0.001) \\ \hdashline
  ts-400-06 & 0.637(0.011) & 0.689(0.012) & 0.773(0.016) & 0.773(0.016) & 0.998(0.002) \\
  ts-400-15 & 0.545(0.008) & 0.598(0.006) & 0.811(0.015) & 0.822(0.014) & 1.000(0.000) \\
  ts-400-30 & 0.534(0.006) & 0.581(0.004) & 0.872(0.014) & 0.884(0.012) & 1.000(0.000) \\
   \hline
\end{tabular}
\end{center}
\end{table*}

\begin{figure*}[!th]
\includegraphics[width=\textwidth]{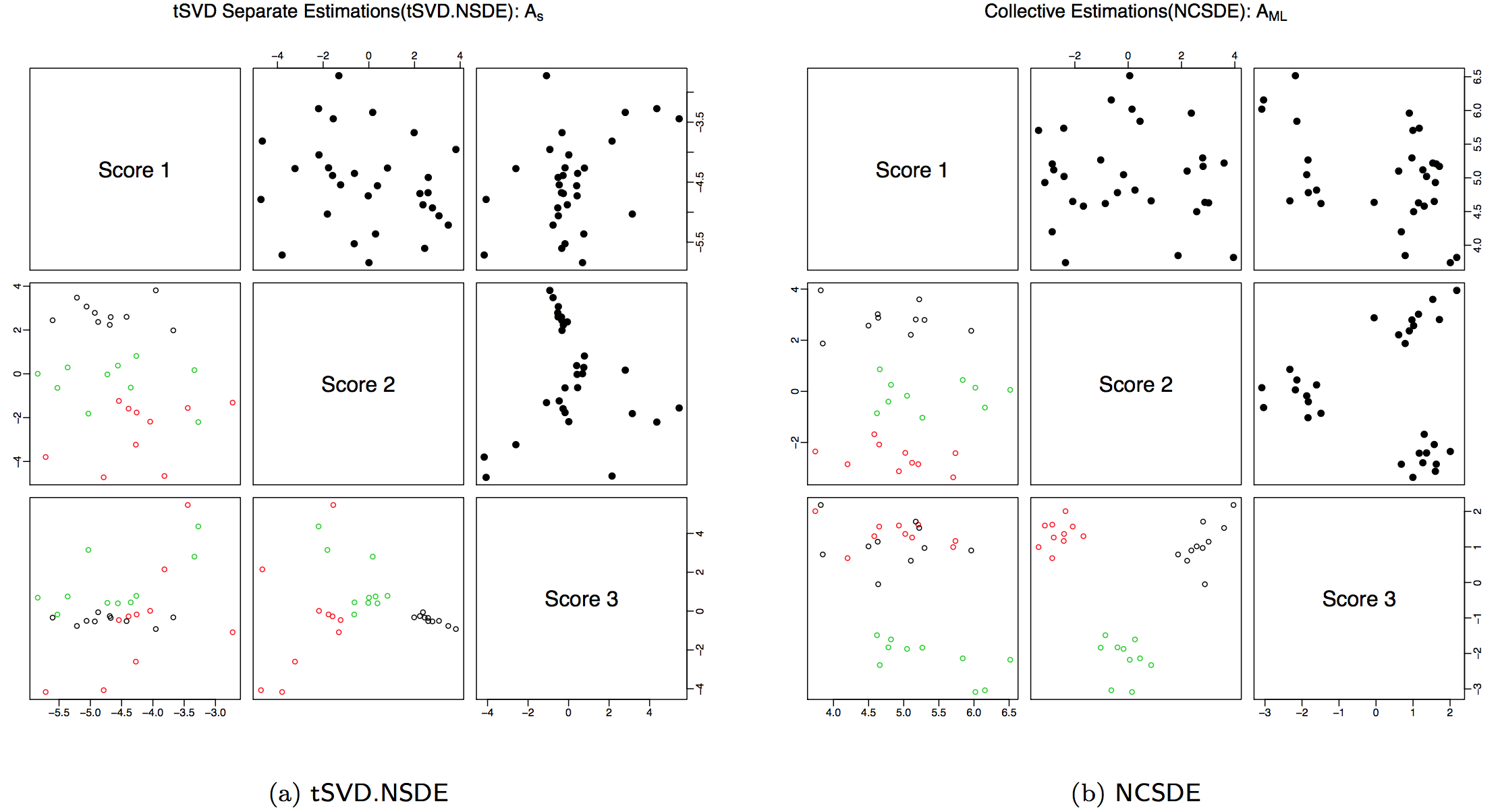}
\caption{A matrix plot that assesses the relationships among several pairs of scores, $\bA$, in \textsf{tSVD.NSDE} ($\bA_s$) and \textsf{NCSDE} ($\widehat{\bA}$), when clustering a randomly selected simulation run with $m=30$ and $n=400$.}
\label{fig3:fig}
\end{figure*}

\noindent We proceeded with a two-step procedure to visualize the level of similarity among the SDFs:
\bi
\item[{\it i})] Obtain the $\tilde{n}\times m$ matrix ($\hat{\bf f}$) from any of the approaches given; then
\item[{\it ii})] Use a multidimensional scaling (MDS) technique \citep{Borg2005}.
\ei
However, another advantage of the proposed method is that the coefficients of the basis expansion for the fitted densities ($\bA$ matrix) provide a low-dimensional representation that can be directly used to visualize and cluster the spectral densities. Figure \ref{fig3:fig} shows the scatter plot of the first three coefficients ($\bA_{.1}$, $\bA_{.2}$, and $\bA_{.3}$) for the last two approaches in subsection \ref{comp.approach} (\textsf{tSVD.NSDE} and \textsf{NCSDE}). We observe a clear segregation into three classes using {\sf NCSDE}, indicating that these coefficients are useful for clustering purposes.

\section{EEG data application}\label{sec:app}
\subsection{EEG data introduction}
\begin{figure}[!ht]
\centering
\includegraphics[width=\textwidth]{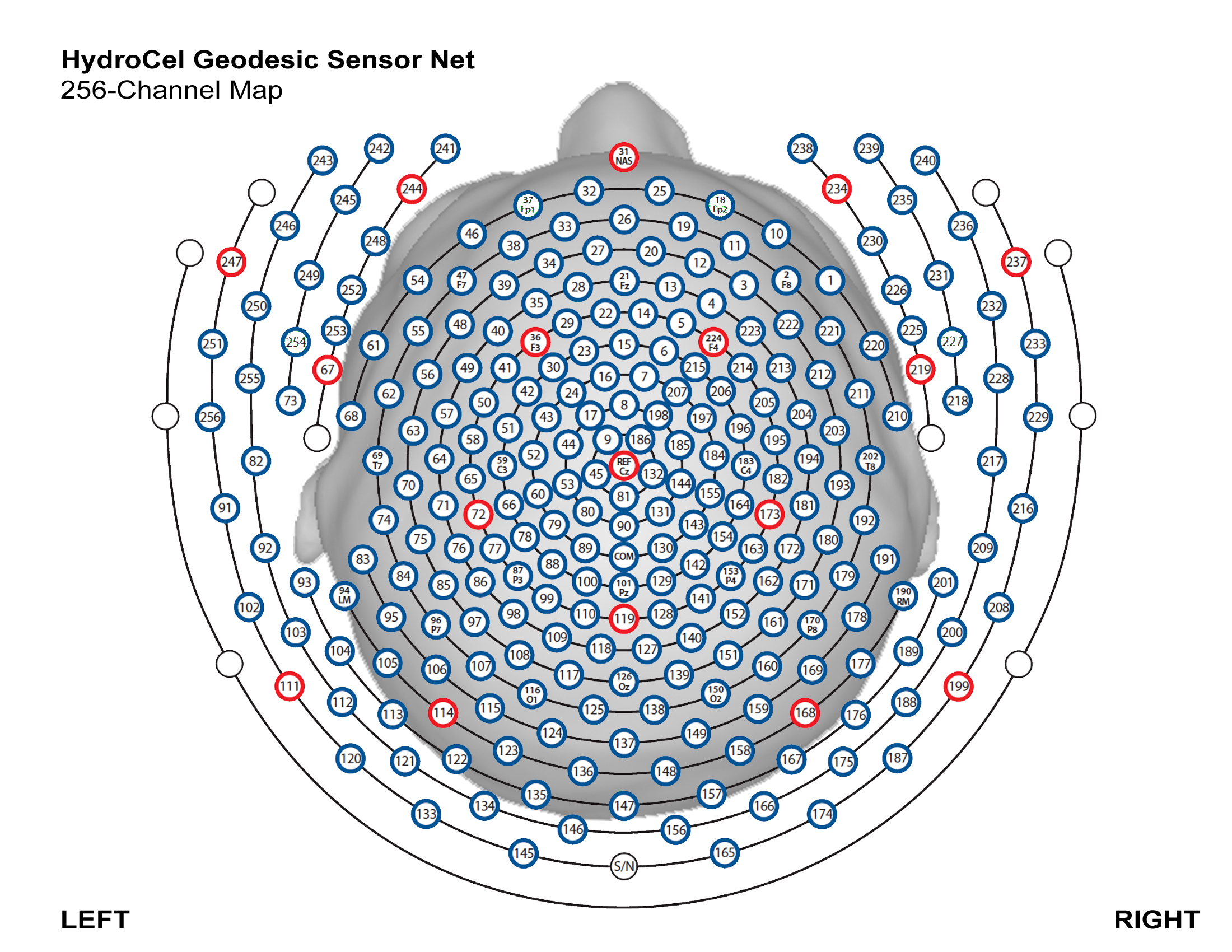}
\caption{The locations of the 256 channels on the scalp surface.}
\label{256}
\end{figure}

The EEG data was collected during the resting state from a single subject, a male student randomly selected from the 17 subjects in the experiment, which is described in detail in \citet{wu2014resting}. During the experiment, the EEG signals were recorded from 256 channels on the scalp surface, with a millisecond resolution (1000 recordings per second). From the 256 channels, 62 channels were eliminated due to issues with the quality of the data \citep{wu2014resting}. The locations of the channels are shown in Figure \ref{256}. In this application, we cluster the remaining 194 channels using the time series data from the first and the last minutes. The total length of the time series is 60,000. However, we only consider the low-frequency bands (i.e., we truncate the perdiodograms at length $3000$).
\subsection{Clustering results}

\begin{figure*}[!th]
\centering
\includegraphics[width=.75\textwidth]{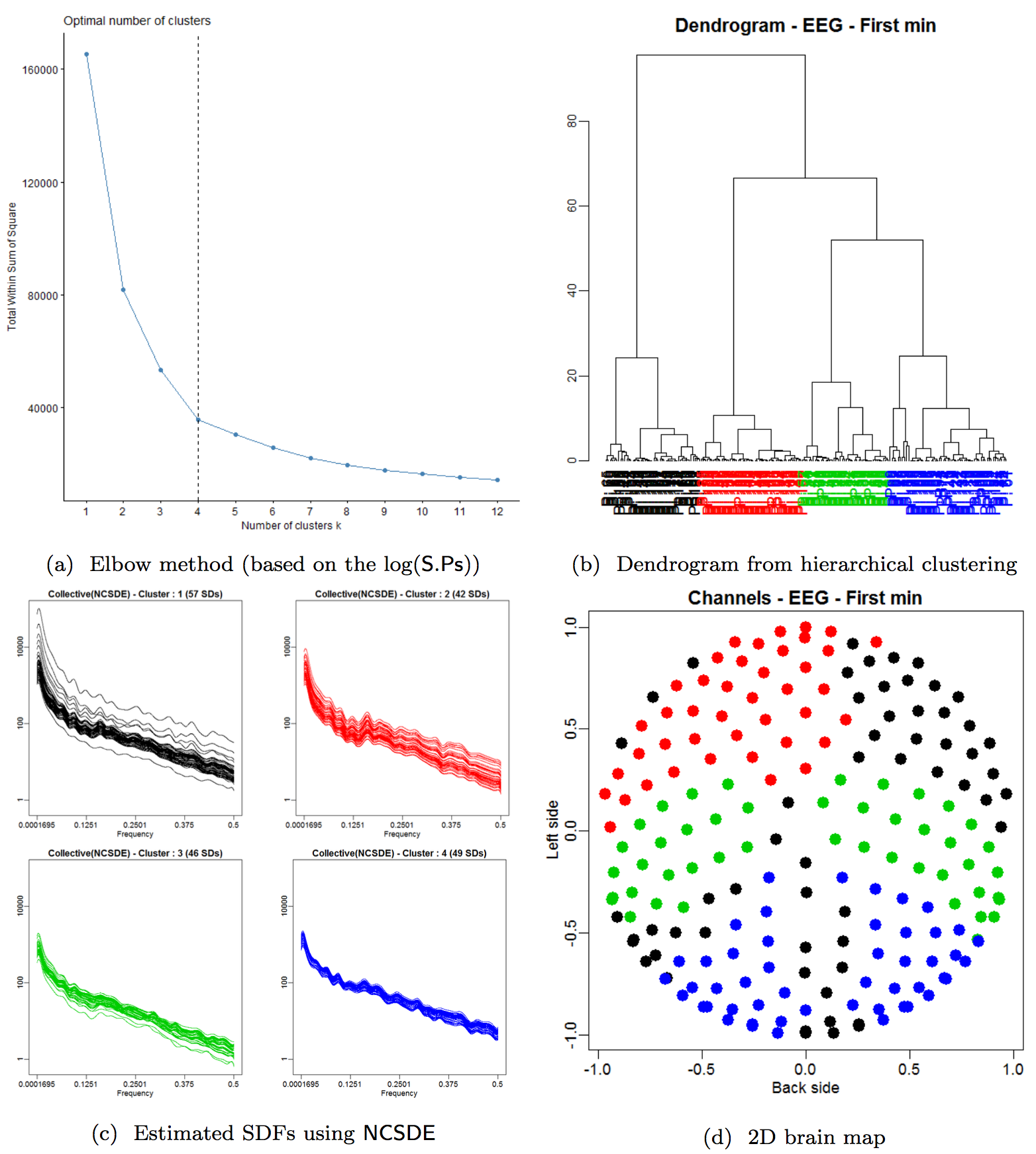}
\caption{The clustering results from the first minute. The top left plot (a) is the result of the elbow method based on the $\log(\mathsf{S.Ps})$ and the top right plot (b) is the result of the hierarchical clustering based on the $\widehat{\bA}$ matrix. The bottom left plots (c) are the estimated SDFs using the \textsf{NCSDE} method, and the last plot (d) is the 2D brain map that illustrates the clustering results from the first minute of data.}
\label{first}
\end{figure*}

\begin{figure*}[!th]
\centering
\includegraphics[width=.75\textwidth]{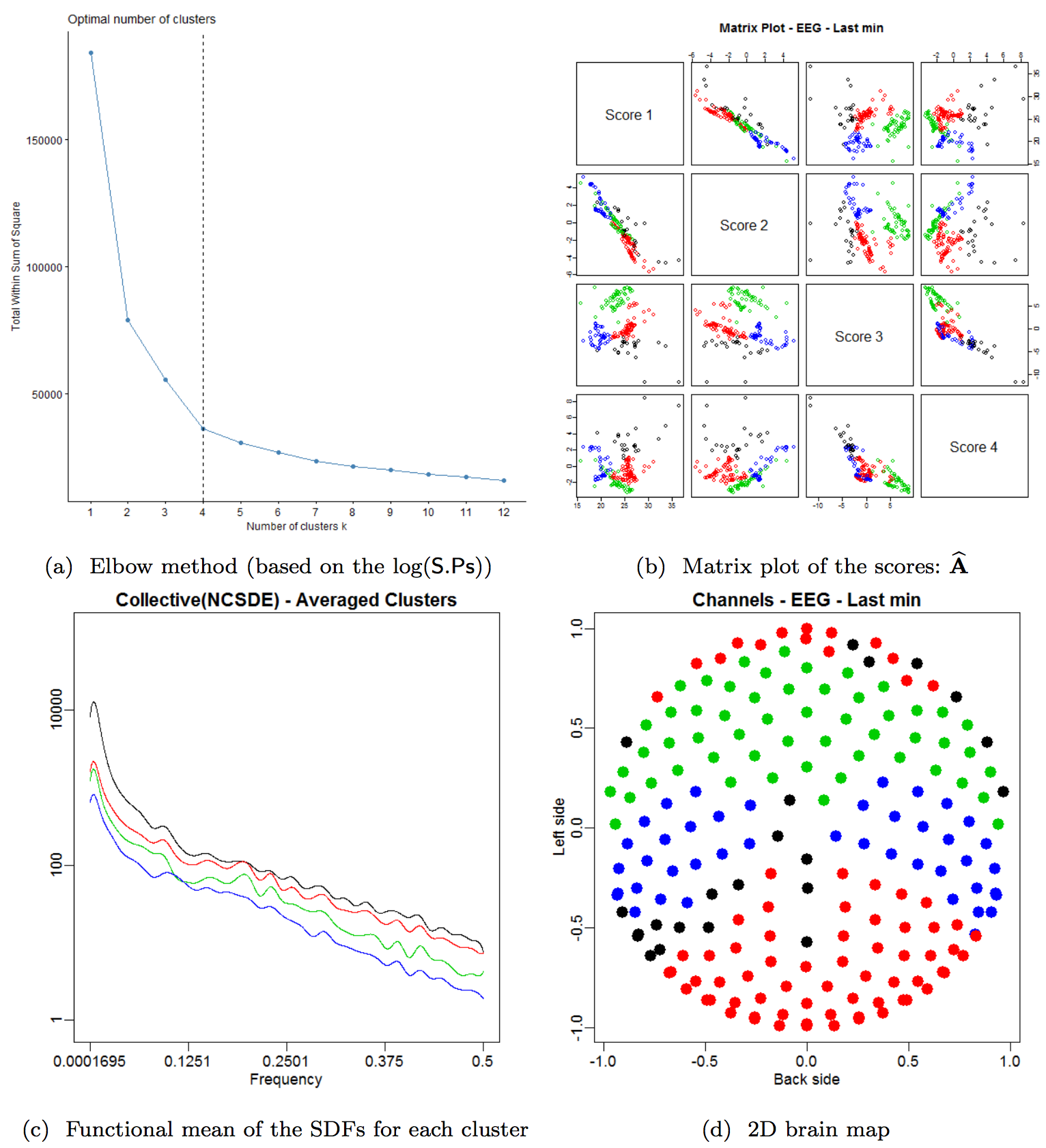}
\caption{The clustering result from the last minute. The top left plot (a) is the result of the elbow method based on the $\log(\mathsf{S.Ps})$ and the top right plot (b) is the result of matrix plot of the scores ($\widehat{\bA}$ matrix). The bottom left plot (c) is the functional mean of the estimated SDFs using the \textsf{NCSDE} method for each cluster, and the last plot (d) is the 2D brain map that illustrates the clustering results from the last minute of data.}
\label{last}
\end{figure*}

The elbow method (Figures \ref{first}(a) and \ref{last}(a) suggested the presence of 4 different clusters in both cases, the first and the last minutes. Therefore, we set $\tilde{k}=4$ for both cases and proceeded with the NCSDE iterative procedure to obtain the Whittle maximum likelihood estimator.

For the data from the first minute, the algorithm converged after $33$ iterations. Figure \ref{first}(b) provides the result of the hierarchical clustering algorithm based on the estimated coefficients of the basis expansion, $\widehat{\bA}$ matrix, where the associated tree has been cut into $\tilde{k}=4$ clusters. Out of the $194$ channels, $57,\ 42,\ 46$, and $49$ SDFs were assigned respectively to the four clusters. Figure \ref{first}(c) shows the $194$ estimated SDFs associated with the four different clusters. Figure \ref{first}(d) presents the clustering results in the 2D brain map.

The results from the last minute of data are presented in Figure \ref{last}. The algorithm converged after $170$ iterations, and Figure \ref{last}(b) provides the hierarchical clustering result based on the estimated $\widehat{\bA}$ matrix for the last minute of data. By cutting the tree into $\tilde{k}=4$ clusters, we determined that $22,\ 75,\ 54$, and $43$ SDFs should be assigned to the first through fourth clusters, respectively. Figure \ref{last}(c) shows the functional means of the estimated SDFs in each cluster to illustrate the difference between the four groups. Figure \ref{last}(d) presents the clustering results in the 2D brain map.

\noindent
As shown in Figure \ref{first} and Figure \ref{last}, we have the following findings:
\begin{enumerate}
\item For both datasets (first and last minutes), the elbow method suggested using four clusters, and both of the brain maps provide relatively well-separated, symmetric regions from the clustering results.
\item All of the brain maps give three regions: the front, middle , and back regions. From the functional mean of each cluster, we find that the channels in the middle region of the brain have a lower density in the low-frequency band than the front and back regions.
\item Both data are associated with the resting state; when we compare the results from the first and last minutes, we find that there are no significant differences between them, except in the front region where we obtain two clusters in first minute (Figure \ref{first}(d)), but just one cluster in last minute (Figure \ref{last}(d)).
\item The most left-front channels in Figure \ref{last}(d) have a different pattern from the other channels.
\end{enumerate}

\section{Conclusion}\label{sec:end}
A novel approach for collectively estimating multiple SDFs was developed in this paper. By pooling data from different time series in the frequency domain and using a shared basis to represent the SDFs, the collective estimation approach is statistically more efficient than non-collective estimation approaches. The proposed method uses the penalized Whittle likelihood approximation to yield a flexible family of spectral densities. As an output of the new method, each estimated log-spectral density is expressed in a basis expansion where the basis is estimated from the data, assuming that the SDFs lie in a low-dimensional manifold of the large space spanned by a pre-specified rich basis. The collective spectral density estimation approach is widely applicable when there is a need to estimate multiple SDFs from different populations. Moreover, the coefficients of the basis expansion for the fitted spectral densities provide a concise, low-dimensional representation that could be useful for visualization and clustering.
Another advantage of the new procedure is that it speeds up the process by updating the smoothness parameter within the Newton-Raphson iterations and avoids a grid search over the space of the smoothing parameter, $\lambda$, which could be very time consuming. 

A web application that can be used by the research community to reproduce the results in this paper or to estimate SDFs collectively based on NCSDE for any other related applications is available at  ``\href{https://ncsde.shinyapps.io/NCSDE}{https://ncsde.shinyapps.io/NCSDE}.''
\end{doublespace}

\bibliography{reff}

\end{document}